\documentclass[a4paper,USenglish,11pt]{article}
\usepackage{complexity_opaque_robots}

\title{Computational Power of Opaque Robots}

\author[1]{\nameFELE}
\author[ ]{Lucia Mambretti}
\author[1]{\nameMERE}
\author[1]{\namePALA}

\affil[1]{\small{\dicoaffiliationstring}}
\affil[]{\{\mailFELE, \mailMERE, \mailPALA\}@unimi.it}
\date{}

\begin{document}

\maketitle

\begin{abstract}
In the field of distributed computing by robot swarms, the research comprehends manifold models where robots operate in the Euclidean plane through a sequence of \emph{look-compute-move} cycles.
Models under study differ for \emph{(i)}~the possibility of storing constant-size information, \emph{(ii)}~the possibility of communicating constant-size information, and \emph{(iii)}~the synchronization mode.
By varying features~\emph{(i,ii)}, we obtain the noted four base models: $\Toblot{}$ (silent and oblivious robots), $\Tfsta{}$ (silent and finite-state robots), $\Tfcom{}$ (oblivious and finite-communication robots), and $\Tlumi{}$ (finite-state and finite-communication robots).
Combining each base model with the three main synchronization modes (\emph{fully synchronous}, \emph{semi-synchronous}, and \emph{asynchronous}), we obtain the well-known 12 models.
Extensive research has studied their \emph{computational power}, proving the hierarchical relations between different models.
However, only \emph{transparent} robots have been considered.

In this work, we study the taxonomy of the 12 models considering \emph{collision-intolerant opaque} robots.
We present six witness problems that prove the majority of the computational relations between the 12 models.
In particular, the last witness problem depicts a peculiar issue occurring in the case of \emph{obstructed visibility} and asynchrony.

\end{abstract}


\section{Introduction}
In the far-ranging field of distributed computing, a significant area concerns \emph{computing by mobile entities}~\cite{survey12, book19}, where tasks are required to be solved by multiple simple and limited entities (also called \emph{robots}) that can move in the environment.
In this realm, manifold theoretical models have been introduced to formalize realistic scenarios (e.g. sensor or drone swarms, dynamic networks, software agents).
One of the most studied is the \emph{look-compute-move} model~\cite{survey12,book19}, where robots, once activated, execute a \emph{cycle} of three steps: they \emph{look} at the environment, they \emph{compute} the next position executing a distributed algorithm, and they \emph{move} to the computed position.

Under the umbrella of the \emph{look-compute-move} macro-model, a vast combination of models has been proposed to formalize different robot capabilities and to study how model settings affect its computational power.
In this respect, robots are assumed to possess very limited and restricted features, in order to find the minimal sets of capabilities which are required to achieve a given task.
Accordingly, robots are assumed to be \emph{autonomous}, \emph{indistinguishable}, \emph{anonymous}, and \emph{homogeneous}: namely, they act without any central control, they cannot distinguish themselves by external appearance or by ids, they possess the same features, they execute the same algorithm in a decentralized way.
Moreover, most of the literature considers \emph{punctiform} robots which cannot communicate with other robots (\emph{silent}), without any persistent memory (\emph{oblivious}), without any agreement on a global coordinate system, or chirality, or a unit measure (\emph{disoriented}).
Besides robot capabilities, different model \emph{environments} have been proposed to study diverse scenarios.
The existing models can be mainly divided into two groups: the models where robots act on the Euclidean plane \cite{ BKAS21,  FMP23,FPS2017,BOOK19_KM}, and the models where robots act on discrete spaces (generally graphs, rings, or lattices) \cite{ DFLMS18, DFN16b, DFPS03,BOOK19_L}.
According to the \emph{synchronization mode}, robots may be synchronized (time is globally divided into rounds) or not.
Specifically, literature proposes three main modes: the \emph{fully synchronous} mode (\fully{}), where all robots execute each step of the \emph{look-compute-move} cycle synchronously in one round, the \emph{semi-synchronous} mode (\semi), where at each round a random subset of robots act synchronously, and the \emph{asynchronous} mode (\async), where robots act without any synchronization assumption.

The traditional problems studied for swarms of mobile entities include \texttt{Pattern Formation} \cite{ BKAS21,FMP18, FMP23, FPS2017,BOOK19_Pattern,SS96,SY99,YS10}, \texttt{Gathering} \cite{ DSKN16,DFLMS18, DFPS03, BOOK19_gathering, KLOTW19}, \texttt{Scattering} \cite{IKBT18, MC22}, \texttt{Flocking} \cite{CP07}.
A common goal of the algorithmic investigation is to reduce the model capabilities required to solve a given problem or to prove the impossibility of solving it under a certain set of capabilities.
This approach has led to describing the \emph{computational power} of a given model (i.e. the set of problems it can solve) and outlining the hierarchical relations (dominance, equivalence, or orthogonality) among different models.
In the last decade, multiple works~\cite{BFK21,DFPSY16,DFN16b,DFN16a,DDFN18,FSSW23} have inspected and compared the computational power of different models which differ in robot features and synchronization mode.
According to the robot features, they have investigated how the communication and storage capabilities affect the computational power of the robots. 
Starting from the classical model where robots are both \emph{oblivious} and \emph{silent} (i.e. without any means of storage or communication), researchers have investigated how the possession of a persistent memory or communication means changes the power of such models.
To characterize these extra properties, they proposed to add a \emph{constant-size light} to each robot which can assume a color chosen among a constant and fixed set of colors. 
Such light is \emph{persistent} (so the color is maintained until the next update), it can be updated at the
beginning of a move step, and it can be internally or externally visible.
Specifically, the literature focuses on four classes of robots: the $\oblot{}$ class, where robots are assumed to be \emph{oblivious} and \emph{silent}, the $\fsta{}$ class, where each robot is embedded with an \emph{internal light} (visible just to the robot, thus providing a persistent memory), the $\fcom{}$ class, where each robot is embedded with an \emph{external light} (visible just to the other robots, thus providing communication means), and the $\lumi{}$ class, where each robot is embedded with an external and internal light.
According to the synchronization mode, each class has been studied under the three settings: \fully, \semi, and \async.

Besides some trivial relations between a pair of models that only differ because the first one enjoys a capability that the second one lacks, 
other model relations may not be obvious to identify.
This is especially true for models characterized by completely different capabilities, so it may be difficult to understand which combination of capabilities is more powerful.
In these cases, the literature has attempted to illustrate some \emph{simulators} to prove the equivalence between models, or some \emph{witness problems} to prove their strict dominance or orthogonality.
Specifically, in~\cite{BFK21,DFPSY16, FSSW23,FSW20}, the authors study the computational power of transparent robots that can move on the Euclidean plane, assuming multiple robots can occupy the same positions (\emph{multiplicity}).
In~\cite{DFN16b,DFN16a,DDFN18}, the authors make the same effort but for robots acting on graphs. 
In~\cite{BFKPSW22}, the authors consider \emph{energy-constraint} robots, i.e. robots that necessitate an idle round to restore the needed energy to perform a new cycle. 

\subparagraph*{Related works and our contributions.}
Our work is inspired by the papers~\cite{BFK21, DFPSY16,FSSW23, FSW20} where the authors exhibit the complete taxonomy of the 12 models of robots that can freely move on the Euclidean plane.
Such models vary for the synchronization mode and for the possibility to memorize and communicate. 
However they are assumed to be transparent, thus always guaranteeing complete visibility for the swarm, and collision-tolerant, thus allowing robots to occupy the same position at the same time.

In this paper, we investigate the computational power of \emph{opaque robots}, i.e. robots that cannot see beyond a collinear robot.
Opaqueness introduces a remarkable difficulty in the design of correct algorithms to solve some classical problems \cite{BKAS21,FMP18,  FMP23}.
In fact, the \emph{obstructed visibility} leads to critical issues to be addressed in the algorithmic strategies: robots may not be aware of the cardinality of the swarm, robots may not be aware if there are some moving robots in the \async\ mode, robots may not know the complete topology of the current configuration, robots may compute the next action based on partial information.
As a matter of fact, \emph{ad hoc} techniques are needed to cope with this visibility limitation \cite{SVT21,SVTBR16}.

Besides the \emph{opaqueness} feature, our model differs from~\cite{BFK21, DFPSY16, FSSW23, FSW20} since robots do not tolerate collisions (so we drop the \emph{multiplicity} assumption).
The reason behind this choice is twofold, and it is coherent with the related literature \cite{BKAS21,FMP18,  FMP23, SVT21,SVTBR16}. 
Firstly, assuming collision intolerance leads to the formalization and analysis of more realistic models, as does assuming robot opaqueness.
Secondly, dropping the multiplicity assumption is coherent with the hypothesis of obstructed visibility in the case of collinearity.
As a matter of fact, a multiplicity of two robots forms a \quotes{degenerate} collinearity with any other robot of the swarm, for which it would be unnatural to state the visibility relation in this special case.
In this respect, some witness problems introduced in~\cite{BFK21, FSW20} cannot be applied under our model, which needs a new study with specific witness problems.

In the first part of this work, we expose a preliminary study of the relations between transparent and opaque models.
Intuitively, a transparent model seems to computationally dominate the same model but with opaque robots.
In \Cref{sec:trasparent_vs_opaque} we formally prove this strict dominance: endowing a model with transparency increases its computational power, allowing it to solve more problems. 
As a consequence, this result highlights that constant-size (internal or external) lights are not always sufficient to compensate for robot obstructed visibility.

In the second part of this work (\Cref{sec:taxonomy}), we present six witness problems showing the majority of the hierarchical relations among models of \emph{collision-intolerant opaque robots}, thus providing a first overview of their computational taxonomy.
For the sake of space, all relations proved in this work will be compactly shown in the theorems in \Cref{sec:taxonomy} (i.e. without splitting them in multiple corollaries). 
See \Cref{appendix} for the proofs of such theorems.

\section{Preliminaries}

\subsection{Models}
This work compares 12 robot models that differ in some features.
We here introduce in detail all the \emph{core features} that such models share, and the \emph{variable features} under study.

\subparagraph*{Core features.}
We investigate swarms of autonomous computational mobile robots, which act in the Euclidean plane $\R^2$.
Robots are \emph{indistinguishable} (they cannot be distinguished by external appearance), \emph{anonymous} (they are not provided with any id), \emph{homogeneous} (they execute the same algorithm), and \emph{punctiform} entities.
We consider \emph{opaque} robots so that in the case of three collinear robots $p,q,r$, the endpoint robots $p,r$ cannot see each other.
We assume robots are in the worst condition about orientation: they are \emph{completely disoriented} so that they do not share a global common coordinate system (i.e. no agreement on origin, axis direction, chirality, or unit distance).
Moreover, we assume that the local coordinate system of any robot may change from one activation to another (\emph{variable disorientation}).

All the robots in the swarm are provided with the same deterministic algorithm, which is executed every time the robot is activated.
At each time, a robot can be either \emph{idle} or \emph{active}, according to the scheduler.
When activated, a robot executes a \emph{Look-Compute-Move} cycle: it takes the snapshot of its visible area (\emph{look}), it executes the algorithm using the sole snapshot as input (\emph{compute}), and it travels straight towards the computed destination (\emph{move}). 
If the destination position is equal to the current one, the robot is said to perform a \emph{null movement}.
After the \emph{move} step, the robot becomes idle again.
We consider \emph{rigid} models, i.e. no adversary can stop the motion of a robot\footnote{In \cite{BFK21, DFPSY16, FSSW23,FSW20}, the authors consider both rigid and non-rigid models. 
In the next model comparisons (transparent vs opaque), we consider only rigid models.}.

We deal with a \emph{collision-intolerant model} meaning that it does not tolerate either multiplicity (i.e. no robot can occupy the same location as another robot at the same time) or overlapping trajectories (robots $r$ and $s$ have overlapping trajectories if \emph{(i)} $r$ is moving from $a$ to $a'$, \emph{(ii)} $s$ is moving from $b$ to $b'$, and \emph{(iii)} the segments $\bar{aa'}$ and $\bar{bb'}$ have points in common). 
We refer to both multiplicity and overlapping trajectories as \emph{collisions}.

\subparagraph*{Variable features.}
Regarding the \emph{memory and communication} features of robots, we consider the four models mainly proposed in the literature.
In the $\oblot{}$ model, robots are assumed to be \emph{oblivious} (i.e. they do not have any persistent memory to store data about past cycles) and \emph{silent} (i.e. they do not have any means to communicate with other robots).
In the $\fsta{}$ model, robots are provided with a persistent \emph{internal light} which can assume a color chosen from a constant-size set. 
Such internal light plays the role of a constant-size persistent memory.
In the $\fcom{}$ model, robots are equipped with a persistent \emph{external light} visible only to other robots, which can assume a color chosen in a constant-size set of colors. 
Indeed, external lights can be exploited by the swarm to communicate some messages to the visible robots.
Lastly, the $\lumi{}$ model gather the features of both $\fsta{}$ and $\fcom{}$.
This model assumes \emph{luminous} robots, which are equipped with a light that can be colored using a constant-size set of colors. 
Such light is both visible to the robot itself (working as an internal state) and visible to the other robots (working as an external communication means).

Regarding the \emph{activation and synchronization} of robots, we consider the three modes mainly studied in the literature.
In the \emph{fully synchronous} mode (\fully), time is split into atomic rounds, within which all robots are activated together and execute their look-compute-move steps completely synchronously.
The \emph{semi-synchronous} mode (\semi) differs from \fully{} just for the fact that at each round a random subset of the swarm is activated.
In the \emph{asynchronous} mode (\async), every robot is activated independently from the others, and every cycle step lasts a finite but unpredictable amount of time.
For the \semi{} and \async{} modes, robots do not know which are the activated robots at each instant. 
Moreover, we always assume the \emph{fairness condition}: for each time $t$ and for each robot $r$, there exists a time $t' > t$ such that $r$ is activated.
This condition allows us to compute time complexity considering the number of \emph{epochs}, where an epoch is a minimal time frame within which each robot is activated at least once.
The selection of the subset of robots activated at every time is made by an adversarial \emph{scheduler}.
Formally, let $\swarm=\{r_1,\dots,r_n\}$ be a swarm of $n$ robots, and let $\T$ be a time domain which could be discrete~$\nat_{\geq 0}$ (in \fully\ and \semi) or continuous~$\R_{\geq 0}$ (in \async). 
An \emph{activation scheduling} is a function $S:\T \to 2^\swarm$ defining the subset of the swarm that is activated at a specific time.

\subparagraph*{Notation.}
We use the notation $\opaque{X}^Y$ to indicate a model for opaque robots that possess all the above core features and that has $X$ as communication-storage setting and $Y$ as synchronization mode, where $X \in \{\Toblot{}, \Tfsta{}, \Tfcom{}, \Tlumi{}\}$ and $Y \in \{\ftt, \stt, \att\}$ (\fully, \semi, \async, resp.).
Consistently with the notation used in \cite{BFK21, DFPSY16, FSSW23,FSW20}, we indicate with $X^Y$ the same model as $\opaque{X}^Y$ but considering transparent robots which tolerate collisions.
We refer to these two classes of models as the \emph{opaque} and \emph{transparent framework}.

\subsection{Problems}
Robot swarms are distributed systems that are aimed at solving problems.
Since in these models robots can just move in the plane, the literature studies problems requiring a swarm to form (a sequence of) geometric patterns, and/or to travel along specific trajectories.
Formally, let us assume a swarm of $n$~robots $\swarm=\{r_1,\dots,r_n\}$ on the Euclidean plane.
When no ambiguity arises, we indicate with $r_i$ both the robot and the point on the plane where $r_i$ is located.
Given an absolute coordinate system $Z$ on $\R^2$, we define the \emph{configuration} of the swarm at time $t$ as the set $C_t=\{(x_1,l_1), \dots, (x_n, l_n)\}$ where $x_i\in\R^2$ is the position of $r_i$ according to $Z$, and $l_i$ is the light color of $r_i$, at time $t$.
In the $\Toblot{}$ model, we always assume $l_i=\col{off}$ for every $r_i\in\swarm$.
A configuration is \emph{valid} if no collision occurs on it.
We define $\C$ as the set of all the valid configurations for $\swarm$.
We say that a configuration $C$ guarantees \emph{complete visibility} if there are no collinearities among robots.

A \emph{problem} $P$ for a swarm of robots is defined\footnote{For our purposes.} as a sequence $(\phi_0,\tau_0,\phi_1,\tau_1,\dots, \phi_m,\tau_m\dots)$ where each $\phi_i$ is a condition on the configuration of the swarm, and where $\tau_i$ is a condition on the intermediate configurations that the swarm is allowed to assume to reach a new configuration for which $\phi_{i+1}$ holds. 
We call such sequence the \emph{request of the problem} $P$.
The initial condition $\phi_0$ must include the clause stating that $l_i=\col{off}$ for every $r_i\in\swarm$.
Except for this clause, since $P$ might be solved without lights and under any synchronization mode, $\phi_i,\tau_i$ must not impose any conditions on light colors or the number of cycles, for each $i$.

Starting from an initial configuration $C_0$ for which $\phi_0$ is true, $P$ is said to \emph{be solved} under a scheduling mode if, for each scheduling under the given mode, there exists an algorithm $\A$ through which the swarm forms a sequence of configurations $(C_1,\dots, C_m,\dots)$ such that $\phi_i$ holds in $C_i$ for each $i\geq 1$, and such that $\tau_{i-1}$ holds during the formation of $C_i$ starting from $C_{i-1}$.
If the request of the problem is finite, the last condition $\tau_m$ requires the swarm to stay still after having satisfied the last condition $\phi_m$ of the request.

Given an initial configuration $C_0$, a scheduling on a time domain $\T$
 and an algorithm $\A$ solving $P$, we define the sequence $\{C(t)\}_{t\in\T}$ as the \emph{evolution} of $\A$, where $C(t)$ is the configuration reached at time $t$ executing $\A$ according to the scheduling.

\subsection{Computational Relations}
Given a model $M$, we indicate with $\prob{M}$ the set of problems solved under $M$, i.e. the \emph{computational power} of $M$.
Given two models $M_1,M_2$, we define the following relations:
\begin{itemize}
	\item $M_1$ is \emph{computationally not less powerful} than $M_2$, formally $M_1\geq M_2$, if $\prob{M_1} \supseteq \prob{M_2}$, i.e any problem solvable in $M_2$ is solvable in $M_1$;
	\item $M_1$ is \emph{computationally more powerful} than $M_2$, formally $M_1> M_2$, if $\prob{M_1} \supset \prob{M_2}$, i.e any problem solvable in $M_2$ is solvable in $M_1$ and there exists a problem solvable in $M_1$ that is not solvable in $M_2$;
	\item $M_1$ is \emph{computationally orthogonal} to $M_2$, formally $M_1 \perp M_2$, if $\prob{M_1} \setminus \prob{M_2} \neq \emptyset$ and $\prob{M_2} \setminus \prob{M_1} \neq \emptyset$, i.e there exists a problem solvable in $M_1$ ($M_2$, resp.) that is not solvable in $M_2$ ($M_1$, resp.);
	\item $M_1$ is \emph{computationally equivalent} to $M_2$, formally $M_1 \equiv M_2$, if $\prob{M_1} = \prob{M_2}$, i.e $M_1$ and $M_2$ solve the same set of problems.
\end{itemize}

The following relations trivially follow from the definitions of the models:
$$
\begin{aligned}[c]
\lumi^Y \geq \fsta^Y \geq \oblot^Y &
\mbox{  and } &
\lumi^Y \geq \fcom^Y \geq \oblot^Y
\end{aligned}
$$
$$X^\ftt \geq X^\stt \geq X^\att$$
where $Y \in \{\ftt, \stt, \att\}$ and $X \in \{\Toblot{},\Tfsta{}, \Tfcom{}, \Tlumi{}\}$.
Indeed, the same relations hold in the opaque framework.

\section{Transparent vs opaque robots}\label{sec:trasparent_vs_opaque}

\begin{theorem}\label{th:notsolv_transp_opaque}
Let $P$ be a problem solved in $\opaque{X}^Y$.
Then $P$ is solved under ${X}^Y$.
\end{theorem}
\begin{proof}
Let $\opaque{\A}$ be an algorithm solving $P$ under $\opaque{X}^Y$.
We can easily construct an algorithm $\A$ solving $P$ under $X^Y$.
Given a robot $r$ and given in input its snapshot $\sigma$ of all the robots, $\A$ computes $\A(\sigma) \sets \opaque{\A}(\opaque{\sigma})$
where $\opaque{\sigma}$ is the snapshot obtained by $\sigma$ removing all the robots which would be hidden from $r$ in case of opaqueness.
$\A$ perfectly simulates $\opaque{\A}$, thus correctly solving $P$ for transparent robots.
\end{proof}

\begin{corollary}\label{cor:dominance_transparent_opaque}
For each $Y \in \{\ftt, \stt, \att\}$ and $X \in \{\Toblot{},\Tfsta{}, \Tfcom{}, \Tlumi{}\}$,
$$\opaque{X}^Y \leq X^Y.$$
\end{corollary}

\begin{problem}[\linestretch]
Let us consider an initial configuration where $n>3$ robots are equally spaced along the same line, say $\gamma$.
Let $d$ be the distance between two adjacent robots.
The problem asks the endpoint robots to move away from their adjacent robot and stop in order to form a new distance $d+\frac{d}{n}$ with them.
They are allowed to travel only along $\gamma$.
The other robots must stay still.
See \Cref{fig:linestrectch}.
\end{problem}

\begin{figure}[!h]
\begin{center}
	\begin{tikzpicture}[scale=0.5, transform shape, font = {\LARGE}]
		\def\d{3.5cm}
		\draw[very thin, dotted] (-3*\d,0) -- (3*\d,0);
		\node (g) at (3.1*\d,0) {$\gamma$};

		\rnode{\colorR}{above}{$$}{\d,0};
		\rnode{\colorR}{above}{$$}{-\d,0};
		\rnode{\colorR}{above}{$$}{0,0};
		\rnode{\colorR}{above}{$$}{2*\d,0};
		\rnode{\colorR}{above}{$$}{-2*\d,0};
		
		\draw[dashed, ->] (2*\d,0) -- (2*\d+1.6*\d*0.2,0);
		\draw[dashed, ->] (-2*\d,0) -- (-2*\d-1.6*\d*0.2,0);
		
		\rnode{\colorE}{above}{}{2*\d+2*\d*0.2,0};
		\rnode{\colorE}{above}{}{-2*\d-2*\d*0.2,0};

	\end{tikzpicture}

\end{center}
\caption{\linestretch.}
\label{fig:linestrectch}
\end{figure}

\begin{lemma}\label{lemma:linestretch_in}
\linestretch\ is solved under $\Toblot{A}$. 
\end{lemma}
\begin{proof}
The problem is solved under the weakest model of the transparent framework.
In fact, the endpoint robots can compute and head to their destination since they can count all the robots and at least two internal robots fix $d$. 
The final configuration is stable.
\end{proof}

\begin{lemma}\label{lemma:linestretch_notin}
\linestretch\ cannot be solved under $\Olumi{F}$.
\end{lemma}
\begin{proof}
The problem cannot be solved under the strongest model of the opaque framework.
Since the $n$ robots are always collinear by request, they cannot count themselves and so the endpoint robots will never accomplish the task.
Moreover, lights would be inefficient for keeping a swarm counter, due to their constant size.
\end{proof}

\begin{theorem}\label{th:strict_dominance_transparent_opaque}
For each $Y \in \{\ftt, \stt, \att\}$ and $X \in \{\Toblot{},\Tfsta{}, \Tfcom{}, \Tlumi{}\}$,
$$\opaque{X}^Y < X^Y.$$
\end{theorem}
\begin{proof}
The result derives by combining \Cref{cor:dominance_transparent_opaque} with \Cref{lemma:linestretch_in} and \Cref{lemma:linestretch_notin}.
In fact, it holds that $\linestretch \in \prob{X^Y}$ whereas  $\linestretch \notin \prob{\opaque{X}^Y}$ for any $X,Y$.
\end{proof}

\begin{theorem}\label{th:cv_nfixed}
Let $P$ be a problem solved by an algorithm $\A$ under $X^Y$ always avoiding collisions, such that $P$ is defined for a swarm with fixed cardinality, say $k$.
If, given any evolution of $\A$, every robot can see $k$ robots, then the problem can be solved even in $\opaque{X}^Y$.
\end{theorem}

\begin{proof}
Since at any activation, each robot is aware it sees the whole swarm, it can compute its next action by executing $\A$.
This computation results in the solution of the problem considering opaque robots.
\end{proof}

\section{Taxonomy of opaque models}\label{sec:taxonomy}
We present our witness problems to prove some strict dominance ($>$) and orthogonality ($\perp$) relations among opaque models.
Thanks to \Cref{th:notsolv_transp_opaque} and \Cref{th:cv_nfixed}, one of the witness problems presented in~\cite{BFK21} can be used to prove some hierarchical relations to hold in our opaque framework too.
However, other witness problems in~\cite{BFK21, FSW20} are not compliant with our collision-intolerant models; thus, we present specific problems that fit our assumptions.

\subsection{Weakness of $\Ooblot{}$}

\begin{problem}[\triangleroundtrip]
Let $C$ be a configuration where 3 robots are placed so that two of them lay on the vertices of an equilateral triangle (let $a$ be the empty vertex), while the third robot lays on the triangle center.
From $C$, the robot in the center has to move to $a$, forming the new configuration $C'$.
Then, robots have to form $C$ again, where $a$ is again the empty vertex.
See \Cref{tab:triangleroundtrip}.
\end{problem}

\begin{table}[!h]
\begin{center}
\begin{tabular}{c|c|c}

	\begin{tikzpicture}[scale=0.4, transform shape, font = {\LARGE}]
		\def\r{\ray}
		\draw [thin, dotted] (0,0) circle (\r);
		\draw[dashed, ->] (0:0) -- (330:\r*0.8);
		
		\rnode{\colorR}{left}{}{90:\r};
		\rnode{\colorR}{left}{}{210:\r};
		\rnode{\colorE}{right}{$a$}{330:\r};
		\rnode{\colorR}{left}{}{0:0};
		
	\end{tikzpicture}
	&
	\begin{tikzpicture}[scale=0.4, transform shape, font = {\LARGE}]
		\def\r{\ray}
		\draw [thin, dotted] (0,0) circle (\r);
		\draw[dashed, <-] (330:\r*0.2) -- (330:\r*0.8);
		
		\rnode{\colorR}{left}{}{90:\r};
		\rnode{\colorR}{left}{}{210:\r};
		\rnode{\colorR}{right}{$a$}{330:\r};
		\rnode{\colorE}{left}{}{0:0};
	\end{tikzpicture}
	&
	\begin{tikzpicture}[scale=0.4, transform shape, font = {\LARGE}]
		\def\r{\ray}
		\draw [thin, dotted] (0,0) circle (\r);

		\rnode{\colorR}{left}{}{90:\r};
		\rnode{\colorR}{left}{}{210:\r};
		\rnode{\colorE}{right}{$a$}{330:\r};
		\rnode{\colorR}{left}{}{0:0};

	\end{tikzpicture}
	
	\\ 
	$C$& $C'$ & $C$
\end{tabular}
\end{center}
\caption{Configurations in \triangleroundtrip.}
\label{tab:triangleroundtrip}
\end{table}

\triangleroundtrip\ is a sub-case of the problem \texttt{N-gon Round-Trip} defined in~\cite{BFK21} (see Definition~1).

\begin{lemma}\label{lemma:triangle_notin}
$\triangleroundtrip \notin \prob{\Ooblot{F}}$.
\end{lemma}
\begin{proof}
The problem has been shown to not belong to $\Toblot{F}$ (see Lemma 3 in \cite{BFK21}).
In fact, using oblivious and silent robots, there is no way to identify the former empty vertex $a$ due to the full symmetry of $C'$.
By the contrapositive of \Cref{th:notsolv_transp_opaque}, the result holds.
\end{proof}

\begin{lemma}\label{lemma:triangle_in}
$\triangleroundtrip \in \left(\prob{\Ofsta{A}} \cap \prob{\Ofcom{A}}\right)$.
\end{lemma}
\begin{proof}
The problem has been shown to be solved in $\Tfsta{A}$ and $\Tfcom{A}$ (see Lemma 4-5 in \cite{BFK21}).
Since in this version of the problem the cardinality of the swarm is fixed and the robots never create collinearities or collisions, we can apply \Cref{th:cv_nfixed} to state that $\triangleroundtrip$ can be solved both in  $\Ofsta{A}$ and $\Ofcom{A}$.
\end{proof}

\begin{theorem}\label{th:oblot_weak}
Given the schedulers $Y_1= \ftt$, $Y_2= \stt$, $Y_3= \att$, it holds
$$\opaque{\fsta}^{Y_i} > \opaque{\oblot}^{\{Y_j\}_{j\geq i}}$$
$$\opaque{\fcom}^{Y_i} > \opaque{\oblot}^{\{Y_j\}_{j\geq i}}$$
$$\opaque{\lumi}^{Y_i} > \opaque{\oblot}^{\{Y_j\}_{j\geq i}}.$$
\end{theorem}

\subsection{Orthogonality between $\Ofsta{}$ and $\Ofcom{}$}

\begin{problem}[\flipflopflip]
Let $p$, $q$ and $r$ be three robots forming a strictly isosceles triangle so that $dist(p,r)=dist(q,r)$.
Let $\gamma$ be the perpendicular bisector to the line segment $\bar{pq}$ passing through the point $b\in \bar{pq}$.
Let $\gamma'$ ($\gamma''$, resp.) be the semi-line of $\gamma$ starting from $b$ and which contains (does not contain, resp.) $r$.
The problem requires $r$ to perpetually perform three subsequent actions (see \Cref{fig:flipflopflip}), in an infinite loop: \emph{(i)} $r$ must reach a point on $\gamma''\setminus\{b\}$; \emph{(ii)} $r$ must reach a different point on $\gamma''$ in order to move away from $p,q$; \emph{(iii)} $r$ must reach a point on $\gamma'\setminus\{b\}$.
The problem requires $r$ to never leave $\gamma$ and to never stop so that $p,q,r$ form an equilateral triangle.
Robots $p,q$ must stay still.
\end{problem}

\begin{table}[!h]
\begin{center}
\begin{tabular}{c|c|c}

	\begin{tikzpicture}[scale=0.5, transform shape, font = {\LARGE}]
		\def\d{3.7cm}
		\def\l{2}
		\draw[very thin, dotted] (-\d,0) -- (\d,0);
		\node (g') at (\d*1.1,0) {$\gamma'$};
		\node (g'') at (-\d*1.1,0) {$\gamma''$};

		\draw[dashed, <-] (-\d*0.2,0) -- (\d*0.7,0);
		\rnode{\colorR}{above}{$r$}{\d*0.7,0};
		\rnode{\colorE}{above}{}{-\d*0.3,0};
		
		\rnode{\colorR}{left}{$p$}{0,\l*0.5};
		\rnode{\colorR}{left}{$q$}{0,-\l*0.5};

	\end{tikzpicture}
	&
	\begin{tikzpicture}[scale=0.5, transform shape, font = {\LARGE}]
		\def\d{3.7cm}
		\draw[very thin, dotted] (-\d,0) -- (\d,0);
		\def\l{2}
		\node (g') at (\d*1.1,0) {$\gamma'$};
		\node (g'') at (-\d*1.1,0) {$\gamma''$};

		\draw[dashed, <-] (-\d*0.7,0) -- (-\d*0.3,0);
		\rnode{\colorR}{above}{$r$}{-\d*0.3,0};
		\rnode{\colorE}{above}{}{-\d*0.8,0};
		
		\rnode{\colorR}{left}{$p$}{0,\l*0.5};
		\rnode{\colorR}{left}{$q$}{0,-\l*0.5};
	\end{tikzpicture}

	&
	\begin{tikzpicture}[scale=0.5, transform shape, font = {\LARGE}]
			\def\d{3.7cm}
		\draw[very thin, dotted] (-\d,0) -- (\d,0);
		\def\l{2}
		\node (g') at (\d*1.1,0) {$\gamma'$};
		\node (g'') at (-\d*1.1,0) {$\gamma''$};

		\draw[dashed, <-] (\d*0.5,0) -- (-\d*0.8,0);
		\rnode{\colorR}{above}{$r$}{-\d*0.8,0};
		\rnode{\colorE}{above}{}{\d*0.6,0};
		
		\rnode{\colorR}{left}{$p$}{0,\l*0.5};
		\rnode{\colorR}{left}{$q$}{0,-\l*0.5};
	\end{tikzpicture}
	\\ 
	\small{First Flip} & \small{Flop} & \small{Second Flip}
\end{tabular}
\end{center}
\caption{Configurations in \flipflopflip.}
\label{fig:flipflopflip}
\end{table}

\begin{lemma}\label{lemma:flipflopflip_in}
	$\flipflopflip \in \left(\prob{\Ofsta{A}}  \cap \prob{\Ofcom{F}}\right)$.
\end{lemma}

\begin{proof}
We solve the problem in these two models using three colors (\col{flip1}, \col{flop} and \col{flip2}), assuming w.l.o.g. all robots start with the color \col{flip1}.
The problem request guarantees that each robot can recognize its role by geometric conditions.
In $\Ofsta{A}$, $r$ moves along $\gamma$ changing its internal color following the perpetual scheme $(\col{flip1}-\col{flop}-\col{flip2})^\infty$, so that at each activation, $r$ knows which is the current action to be performed.
The robots $p,q$ do not need to change their colors.
In the $\Ofcom{F}$ model, all the robots synchronously update their external colors following the above scheme, so that at each round each robot knows what actions (color setting and move step) have to be accomplished.	
\end{proof}

\begin{lemma}\label{lemma:flipflopflip_notin}
$\flipflopflip \notin \left(\prob{\Ooblot{F}} \cup \prob{\Ofcom{S}}\right)$.
\end{lemma}

\begin{proof}
\flipflopflip\ cannot be solved under an $\Ooblot{}$ model since $r$ would not have any means to understand which movement it has to perform.
Indeed, any strategy encoding the action of $r$ into the distances with $p,q$ fails.
Suppose for example to use $u=dist(p,q)$ as a fixed measure unit, and let $k,h$ be two fixed values, with $0 < k < h $.
Suppose the algorithm implements this strategy: if $dist(p,r) < ku$, then $r$ must execute the first flip, traveling to a position $r'\in \gamma''$ such that $ku\leq dist(p,r') <hu $. 
Otherwise, if $ku\leq dist(p,r) <hu$, then $r$ must execute the flop, moving to a position $r'\in\gamma''$ such that $dist(p,r') \geq hu $.
Lastly,  if $dist(p,r) \geq hu$, then $r$ must execute the second flip, moving to a position $r'\in \gamma'$ such that $dist(p,r') <ku $.
Yet, since $r$ could be placed at any position on $\gamma'$ in the initial configuration, any distance encoding results inefficient for the solution of the problem.

\flipflopflip\ cannot be solved under the $\Ofcom{S}$ model too.
By contradiction, suppose that the problem is solved by a certain algorithm $\A$. 
Let $S$ be a \semi{} activation scheduling under which $\A$ solves the problem.
We show that there exists a \semi{} activation scheduling $S'$ such that \flipflopflip\ is not solved by $\A$. 
Let $t$ be the first round in $S$ where $r$ executes the first flip.
Let $S'$ be a scheduling such that $S'(t') = S(t')$,  $\forall t'\leq t$. 
Clearly, $r$ executes its first flip at the $t$-th round under $S'$.
Suppose that, in the $(t + 1)$-th activation round under $S'$, $r$ is the only one that gets activated, namely $S'(t + 1) = \{r\}$. 
Yet, $r$ has no memory of the previous activation rounds.
As a consequence, $r$ makes again a flip.
Contradiction.
	
\end{proof}

\begin{theorem}\label{th:lumi_strong}
$$\Olumi{A} > \Ofcom{A}$$
$$\Olumi{S} > \Ofcom{S,A}$$
$$\Olumi{F} > \Ofcom{S,A}$$
$$\Ofcom{F} > \Ofcom{S,A}.$$
\end{theorem}

\begin{problem}[\newcomer]
Consider $n+2$ robots, with $n \geq 7$.
Let $n$ robots be placed on the same circle whose ray length is $\rho$.
Let $c$ be a robot lying in the center of the circle.
Let $s$ be a robot external to the circle so that $s$ can see $c$.
The problem requires sequentially forming two configurations. 
First, $s$ must travel along the line $\bar{sc}$ and stop on the boundary of the circle.
Second, $c$ must travel along the radius defined by $s$ and stop in a position $c'$ so that $dist(s,c') = \frac{1}{2}\rho$.
All the other robots must stay still.
See \Cref{fig:newcomer}.
\end{problem}

\begin{table}[!h]
\begin{center}
\begin{tabular}{c|c|c}

	\begin{tikzpicture}[scale=0.4, transform shape, font = {\LARGE}]
		\def\r{\ray}
		\draw [thin, dotted] (0,0) circle (\r);
		\draw[dashed, ->] (180:\r*2) -- (180:\r*1.2);
		
		\rnode{\colorR}{above}{$c$}{0,0};

		\rnode{\colorR}{left}{}{0:\r};
		\rnode{\colorR}{right}{}{280:\r};
		\rnode{\colorR}{left}{}{260:\r};
		\rnode{\colorR}{left}{}{60:\r};
		\rnode{\colorR}{left}{}{100:\r};
		\rnode{\colorR}{right}{}{154:\r};
		\rnode{\colorE}{above left}{}{180:\r};
		\rnode{\colorR}{left}{}{212:\r};

		\rnode{\colorR}{left}{$s$}{180:\r*2};
		
	\end{tikzpicture}
	&
	\begin{tikzpicture}[scale=0.4, transform shape, font = {\LARGE}]
		\def\r{\ray}
		\draw [thin, dotted] (0,0) circle (\r);

		\rnode{\colorR}{above}{$c$}{0,0};

		\rnode{\colorR}{left}{}{0:\r};
		\rnode{\colorR}{right}{}{280:\r};
		\rnode{\colorR}{left}{}{260:\r};
		\rnode{\colorR}{left}{}{60:\r};
		\rnode{\colorR}{left}{}{100:\r};
		\rnode{\colorR}{right}{}{154:\r};
		\rnode{\colorR}{left}{$s$}{180:\r};
		\rnode{\colorR}{left}{}{212:\r};

	\end{tikzpicture}
	&
	\begin{tikzpicture}[scale=0.4, transform shape, font = {\LARGE}]
		\def\r{\ray}
		\draw [thin, dotted] (0,0) circle (\r);
		\draw[dashed, ->] (0,0) -- (-\r*0.4,0);
		
		\rnode{\colorR}{above}{$c$}{0,0};
		\rnode{\colorE}{left}{}{-\r*0.5,0};

		\rnode{\colorR}{left}{}{0:\r};
		\rnode{\colorR}{right}{}{280:\r};
		\rnode{\colorR}{left}{}{260:\r};
		\rnode{\colorR}{left}{}{60:\r};
		\rnode{\colorR}{left}{}{100:\r};
		\rnode{\colorR}{right}{}{154:\r};
		\rnode{\colorR}{left}{$s$}{180:\r};
		\rnode{\colorR}{left}{}{212:\r};

	\end{tikzpicture}
	\\ 
	\small{First Configuration (a)} & \small{Second Configuration (b)} & \small{Third Configuration (c)}
\end{tabular}
\end{center}
\caption{Configurations in \newcomer.}
\label{fig:newcomer}
\end{table}

\begin{lemma}\label{lemma:newcomer_notin}
$\newcomer \notin \prob{\Ofsta{F}}$.
\end{lemma}

\begin{proof}
The impossibility of solving the problem with just internal lights derives from the fact that starting from the second configuration (see \Cref{fig:newcomer}.b) $c$ has no way to recognize which robot is $s$.
Since $s$ can be anywhere in the disposition of the $n+1$ robots on the circle, a constant set of colors would not be sufficient to store robot indices.
\end{proof}

\begin{lemma}\label{lemma:newcomer_in}
$\newcomer \in \prob{\Ofcom{A}}$.
\end{lemma}

\begin{proof}

We show a possible $\Ofcom{A}$ algorithm solving \newcomer\ with two colors: \col{off} and \col{s}.
All the robots are initially set to color \col{off}. 
Each robot can determine its role by the geometry of the configurations ($c$ sees $n\geq 7$ robots equidistant from itself and an external robot, $s$ sees at least four robots forming a circle with a robot on its center, while the other robots can see they lay on a circle with at least other $n-2\geq 5$ robots).
When $s$ is activated, it sets its light to \col{s} and starts to move.
This color is maintained also in its next activations.
When $c$ is activated, if it sees a robot \col{s} on the circle, it can compute its destination correctly.
The last configuration is stable: no other robot will move.
\end{proof}

\begin{theorem}\label{th:fsta_lumi}
Given the schedulers $Y_1= \ftt$, $Y_2= \stt$, $Y_3= \att$, 
$$\opaque{\lumi}^{Y_i} > \opaque{\fsta}^{\{Y_j\}_{j\geq i}}.$$
\end{theorem}

\begin{theorem}\label{th:fsta_fcom}
$$\Ofsta{F,S,A} \perp \Ofcom{S,A}.$$
\end{theorem}

\subsection{Power of \fully}
\begin{problem}[\spinning]
The problem is defined recursively, without any stop conditions.
Consider a configuration $C$ where $n \geq 5$ robots $\{r_0, \dots , r_{n-1}\}$ are located on a circle centered in $O$.
Let  $a_0,\dots, a_{n-1}$ be the related positions of the robots such that it is possible to establish a global clockwise direction (e.g. the one going from $a_0$ to $a_2$, passing through $a_1$). 
Let $\alpha$ be the angle $a_0\hat{O}a_1$, which is the minimum angle in $\{a_i\hat{O}a_{{i+1}}\}_{0\leq i \leq n-1}$.
The problem requires the given configuration to form a new configuration $C'$ by rotating each $r_i$ from $a_i$ to $a'_i$  of an angle $\frac{\alpha}{2}$, following the agreed clockwise direction.
Robots are required only stop on the target points lying on the circumference.
Recursively, the problem demands the same request starting from $C'$.
See \Cref{fig:spinning}.
\end{problem}

\begin{table}[!h]
\begin{center}
\begin{tabular}{c|c|c}

	\begin{tikzpicture}[scale=0.4, transform shape, font = {\LARGE}]
		\def\r{\ray}
		\draw [thin, dotted] (0,0) circle (\r);
		\draw [thin, dotted] (0,0) -- (0:\r);
		\draw [thin, dotted] (0,0) -- (20:\r);
		\node (a) at (10:\r*0.6) {$\alpha$};

		\rnode{\colorR}{right}{$a_0$}{0:\r};
		\rnode{\colorR}{right}{$a_1$}{20:\r};
		\rnode{\colorR}{above right}{$a_2$}{60:\r};
		\rnode{\colorR}{above}{$a_3$}{100:\r};
		\rnode{\colorR}{above left}{$a_4$}{200:\r};
		\rnode{\colorR}{below}{$a_5$}{280:\r};

	\end{tikzpicture}
	&
	\begin{tikzpicture}[scale=0.4, transform shape, font = {\LARGE}]
		\def\r{\ray}
		\draw [thin, dotted] (0,0) circle (\r);
			
		\rnode{\colorR}{right}{}{0:\r};
		\rnode{\colorR}{right}{}{20:\r};
		\rnode{\colorR}{above right}{}{60:\r};
		\rnode{\colorR}{above}{}{100:\r};
		\rnode{\colorR}{below left}{}{200:\r};
		\rnode{\colorR}{below}{}{280:\r};

		\rnode{\colorE}{right}{}{10:\r};
		\rnode{\colorE}{right}{}{30:\r};
		\rnode{\colorE}{above right}{}{70:\r};
		\rnode{\colorE}{left}{}{110:\r};
		\rnode{\colorE}{below left}{}{210:\r};
		\rnode{\colorE}{below}{}{290:\r};

		\def\bea{60}
		\def\ro{6}
		\draw [bend right=\bea,->] (0:\r+\ro) to (10:\r+\ro);
		\draw [bend right=\bea,->] (20:\r+\ro) to (30:\r+\ro);
		\draw [bend right=\bea,->] (60:\r+\ro) to (70:\r+\ro);
		\draw [bend right=\bea,->] (100:\r+\ro) to (110:\r+\ro);
		\draw [bend right=\bea,->] (200:\r+\ro) to (210:\r+\ro);		
		\draw [bend right=\bea,->] (280:\r+\ro) to (290:\r+\ro);

   		\draw [dashed,domain=0:270, ->] plot ({cos(\x)}, {sin(\x)});

	\end{tikzpicture}
	&
	\begin{tikzpicture}[scale=0.4, transform shape, font = {\LARGE}]
		\def\r{\ray}
		\draw [thin, dotted] (0,0) circle (\r);
		\draw [thin, dotted] (0,0) -- (10:\r);
		\draw [thin, dotted] (0,0) -- (30:\r);
		\node (a) at (20:\r*0.6) {$\alpha$};
			
		\rnode{\colorR}{right}{$a'_0$}{10:\r};
		\rnode{\colorR}{above right}{$a'_1$}{30:\r};
		\rnode{\colorR}{above right}{$a'_2$}{70:\r};
		\rnode{\colorR}{above left}{$a'_3$}{110:\r};
		\rnode{\colorR}{below left}{$a'_4$}{200:\r};
		\rnode{\colorR}{below right}{$a'_5$}{290:\r};
	\end{tikzpicture}
	
\end{tabular}
\end{center}
\caption{Configurations in \spinning.}
\label{fig:spinning}
\end{table}

\begin{lemma}\label{lemma:spinning_in}
$\spinning \in \left(\prob{\Ooblot{F}} \cap \prob{\Olumi{A}}\right)$.
\end{lemma}

\begin{proof}
The problem is solvable in $\Ooblot{F}$: each robot always has complete visibility of the swarm, so it is able to determine the rotation center and the rotation angle.
The \fully\ mode guarantees that all the robots agree on the same rotation-angle, at each round.

The problem is solvable under $\Olumi{A}$, by using these colors: \col{off, a0, a1, moving0, moving1, m0, m1, moving, moved, end}.
The algorithm solving the problem executes the same sub-routine perpetually.
This sub-routine implements a complete circle rotation of the swarm.
At the beginning of each circle rotation, all robots are \col{off}.
In the first epoch, the robots $r_0$ and $r_1$ set their lights as \col{a0} and \col{a1}, respectively.
After this setting, robot \col{a0} (\col{a1}, resp.) computes its destination position, sets its light to \col{moving0} (\col{moving1}, resp.) and starts moving.
If a robot $r$, which is not \col{moving0} or \col{moving1} colored, sees a \col{moving0} or \col{moving1} robot, $r$ does nothing.
When a \col{moving0} (\col{moving1}, resp.) robot is activated, it just updates its light to \col{m0} (\col{m1}, resp.).
Once the rotation angle through \col{m0} and \col{m1} has been fixed, the other robots can start their rotation.
If an \col{off} robot $r$ sees both \col{m0} and \col{m1} on the circle, it sets its light as \col{moving} and starts its rotation.
When a \col{moving} robot is activated, it sets its light to \col{moved}.
When a robot sees only \col{m0}, \col{m1}, \col{moved}, or \col{end} robots, then it updates its color to \col{end}.
In the last phase of the sub-routine, if an \col{end} robot can see only \col{end} or \col{off} robots, it resets its color to \col{off}.
Once all robots are \col{off}, the circle rotation is ready to restart.

\end{proof}

\begin{lemma}\label{lemma:spinning_notin}
$\spinning \notin \left(\prob{\Ofsta{S}} \cup \prob{\Ofcom{S}}\right)$.
\end{lemma}

\begin{proof}
\spinning\ is not solvable under $\Ofsta{S}$ since an activated robot $r$ cannot know what movements other robots have already made, thus it cannot determine the rotation-angle.

\spinning\ is not even solvable under model $\Ofcom{S}$. 
Suppose that, by contradiction, there exists an algorithm $\A$ solving \spinning. 
In particular, the problem is solved under an activation scheduler $S$. 
Let $r_0$ be the robot in position $a_0$. 
Let $t_1$ be the activation time, under $S$, of the first round during which $r_0$ performs a non-null movement. 
Let $S'$ be another scheduling, such that
$$S'(t) := S(t)\;\; \forall t < t_1 \mbox{ and } S'(t_1) = S'(t_1 + 1) := \{r_0\}$$
If $\A$ is executed under $S'$, then the execution is the same as $S$ until time $t_1-1$. 
At time $t_1$, robot $r_0$ behaves in the same way as it did under scheduling $S$ but, as no other robot has been activated, then there is no way to keep track of the fact that $r_0$ has already moved. 
At time $t_1+1$ robot $r_0$ is activated again but it cannot understand from geometric conditions that it must stay still. Contradiction.

\end{proof}

\begin{theorem}\label{th:fully_strong}
$$\Ooblot{F} > \Ooblot{S,A}$$
$$\Ofsta{F} > \Ofsta{S,A}$$
$$\Ofcom{F} > \Ofcom{S,A}$$
$$\Ooblot{F} \perp \Ofcom{S,A}$$
$$\Ooblot{F} \perp \Ofsta{S,A}.$$
\end{theorem}

\begin{problem}[\angleshift]
Consider an initial configuration with three robots forming an acute and scalene triangle.
Let $a,b,c$ be the three robots, where $a$ is placed on the greatest angle, say $\alpha$, whereas $c$ is placed on the smallest angle.
Fixing $a$ as the rotation center and following the direction given by $a,b,c$, the problem requires $b$ to rotate of $\alpha$ and $c$ to rotate of $\pi-\alpha$.
The robots are not allowed to stop anywhere else on the plane.
Afterwards, the robots must stay still. 
See \Cref{fig:angleshift}.
\end{problem}

\begin{table}[!h]
\begin{center}
\begin{tabular}{c|c|c}

		\begin{tikzpicture}[scale=0.5, transform shape, font = {\LARGE}]
			\def\b{3cm}
			\def\c{2cm}
			\def\al{80}
			
			\node (a) at (\al*0.5:\c*0.3) {$\alpha$};
		
			\rnode{\colorR}{below}{$a$}{0,0};
			\rnode{\colorR}{below}{$b$}{0:\c};
			\rnode{\colorR }{above}{$c$}{\al:\b};
			
			\draw[] (0,0) -- (0:\c);
			\draw[] (0,0) -- (\al:\b);
			\draw[] (0:\c) -- (\al:\b);
			
		\end{tikzpicture}
	
&
		\begin{tikzpicture}[scale=0.5, transform shape, font = {\LARGE}]
			\def\b{3cm}
			\def\c{2cm}
			\def\al{80}
		
			\rnode{\colorR}{below}{$a$}{0,0};
			\rnode{\colorR}{below}{$b$}{0:\c};
			\rnode{\colorR }{above}{$c$}{\al:\b};
			
			\rnode{\colorE}{above}{$$}{\al:\c};
			\rnode{\colorE}{right}{$$}{180:\b};
			
			\draw[] (0,0) -- (0:\c);
			\draw[] (0,0) -- (\al:\b);
			\draw[] (0:\c) -- (\al:\b);

			\draw[dashed, ->] (\al:\b) -- (180-5:\b);
			\draw[dashed, ->] (0:\c) -- (\al-5:\c);

			\end{tikzpicture}
	&
		\begin{tikzpicture}[scale=0.5, transform shape, font = {\LARGE}]
			\def\b{3cm}
			\def\c{2cm}
			\def\al{70}
		
			\rnode{\colorR}{below}{$a$}{0,0};
			\rnode{\colorR}{above}{$b$}{\al:\c};
			\rnode{\colorR }{below}{$c$}{180:\b};
			
			\draw[] (0,0) -- (\al:\c);
			\draw[] (0,0) -- (180:\b);
			\draw[] (\al:\c) -- (180:\b);
		\end{tikzpicture}
		\\
		Initial configuration.&Required movements. & Final configuration.

	\end{tabular}
\end{center}
\caption{\angleshift.}
\label{fig:angleshift}
\end{table}

\begin{lemma}\label{lemma:angleshift_innotin}
	$\angleshift \in \left(\prob{\Ooblot{F}}\setminus \prob{\Olumi{S}}\right)$.
\end{lemma}
\begin{proof}
	
\angleshift\ is solvable under any \fully\ model: if $b$ and $c$ perform their cycles at the same time, then they correctly compute their target position.
The final configuration is stable since it always forms an obtuse triangle (terminal condition).

Instead, the swarm can suffer from information loss in \semi, making \angleshift\ unsolvable even under $\Olumi{S}$.
In fact, suppose that in the initial configuration only $b$ is activated.
After $b$'s movement, the three robots turn out to be collinear in the reached configuration. 
As a result, $c$ has no means to recompute $\alpha$, whether $c$ uses the geometry of the configuration or uses constant-size lights.
The same happens even if only $c$ is activated.
\end{proof}

\begin{theorem}\label{th_lumif}
$$\Olumi{F} > \Olumi{S,A}$$
$$\Ooblot{F} \perp \Olumi{S,A}$$
$$\Ofsta{F} \perp \Olumi{S,A}.$$
\end{theorem}

\subsection{Opaqueness and asynchrony}

We now introduce the \pseudo\ problem which shows a peculiar issue occurring in case of obstructed visibility and asynchrony.

\begin{definition}
	Given a regular $n$-gon $\mathcal{N}$, for any $n\geq 4$, a \emph{pseudo-polygon} $\Q$ is a subset of vertices of $\mathcal{N}$, such that $|\mathcal{Q}|\geq \frac{n}{2} +1 $. 
	We call $\mathcal{N}$ the \emph{associated polygon} with respect to $\mathcal{Q}$.
\end{definition}

	Given a pseudo-polygon $\mathcal{Q}$, it is always possible to determine the associated polygon, which is unique. 
	In fact, as $\mathcal{Q}$ contains at least three vertices, the circumscribed circle is univocally defined.
	Moreover, since $\mathcal{Q}$ contains more than half of the vertices of the associated $n$-gon, there always exist at least two vertices that are adjacent in $\mathcal{N}$.
	So, it is always possible to univocally establish the associated polygon from a pseudo-polygon.

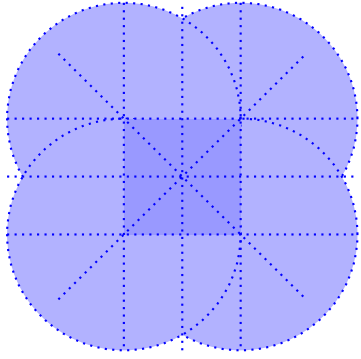
\begin{figure}[h!]
\centering
	\begin{tikzpicture}[scale=0.4, transform shape, font = {\LARGE}]
		\def\l{\ray*0.6}
		\def\n{4}
		\def\a{360/\n}
		\def\o{3}
		\def\c{blue}
		\def\opa{30}

		\draw [thick, dotted, \c, fill=\c!\opa, opacity=0.4] (\l,\l) circle (\l*2);
		\draw[thick, dotted, \c, fill=\c!\opa, opacity=0.4] (\l,-\l) circle (\l*2);
		\draw [thick, dotted, \c, fill=\c!\opa, opacity=0.4](-\l,\l) circle (\l*2);
		\draw [thick, dotted, \c, fill=\c!\opa, opacity=0.4] (-\l,-\l) circle (\l*2);

		\fill[\c!40!white] (\l,\l) rectangle (-\l,-\l);

		\draw [thick, dotted, \c] (\l,\l*\o) -- (\l,-\l*\o);
		\draw [thick, dotted, \c] (-\l*\o,\l) -- (\l*\o,\l);
		\draw [thick, dotted, \c] (-\l,\l*\o) -- (-\l,-\l*\o);
		\draw [thick, dotted, \c] (-\l*\o,-\l) -- (\l*\o,-\l);

		\draw [thick, dotted, \c] ({-\l*\o*cos(45)},{\l*\o*cos(45)}) -- ({\l*\o*cos(45},{-\l*\o*cos(45)});
		\draw [thick, dotted, \c] ({-\l*\o*cos(45)},{-\l*\o*cos(45)}) -- ({\l*\o*cos(45},{\l*\o*cos(45)});

		\draw [thick, dotted, \c] (-\l*\o,0) -- (\l*\o,0);
		\draw [thick, dotted, \c] (0,-\l*\o) -- (0,\l*\o);

	\end{tikzpicture}
\caption{The safe zone of the square comprehends all the points not belonging to the blue-colored (infinite) lines and zones.}
\label{fig:safe_zone}
\end{figure}

\begin{definition}
	A \emph{safe zone} of a regular polygon is the locus of all points $x$ in the plane such that:
	\begin{itemize}[noitemsep]
		\item $x$ is external to the regular polygon;
		\item $x$ is not aligned with any of the two vertices of the associated polygon;
		\item $x$ does not lie on the bisector of any edge of the associated polygon (equivalently, $x$ is not equally distanced from any two adjacent vertices);
		\item if $\ell$ is the length of the edge of the polygon, then the distance between $x$ and any vertex of the polygon is at least $\ell$.
	\end{itemize}
\end{definition}
\Cref{fig:safe_zone} depicts the (complement of the) safe zone of a square.

\begin{figure}[h!]
\centering
	\begin{tikzpicture}[scale=0.4, transform shape, font = {\LARGE}, extended line/.style={shorten >=-#1,shorten <=-#1*0.3},  extended line/.default=1cm]
		\def\r{\ray}
		\def\n{8}
		\def\a{360/\n}
		\draw [thin, dotted] (0,0) circle (\r);
		
		\foreach \i in {3}
			{\rnode{\colorE}{left}{}{\a*0.5+\i*\a:\r}; 
		}

		\foreach \i/\x/\p in {0/c/below left,1/a/above,2/b/above}
			{\rnode{\colorR}{\p}{$\x$}{\a*0.5+\i*\a:\r}; 
		}

		\foreach \i/\x in {4,5,6,7}
			{\rnode{\colorR}{left}{}{\a*0.5+\i*\a:\r}; 
		}

		\draw [thin, dotted, blue, extended line=3cm] (2*\a+\a*0.5:\r) -- ({\r*cos(20)}, {\r*sin(20)});

		\rnode{\colorR}{left}{$w$}{265:\r*2.3}
		\rnode{\colorE}{below}{$x$}{-8:\r*1.9}
		\draw [thin, dashed, ->] (\a*0.5+\a:\r) -- (-6.3:\r*1.9);

	\end{tikzpicture}
\caption{The \pseudo\ problem associated with an octagon.}
\label{fig:pseudo}
\end{figure}
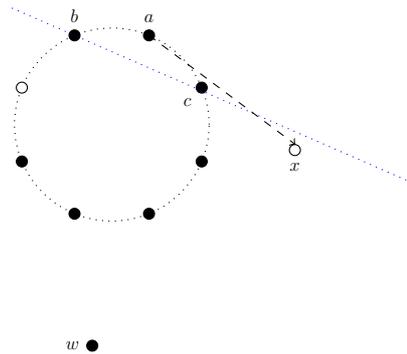

\begin{problem}[\pseudo]\label{prob:pseudo}
Let $\mathcal{N}$ be a regular $n$-gon with $n\geq 6$ vertices.
Let $\Q$ be a pseudo-polygon of $m\geq \frac{n}{2} +2$ vertices, associated with $\mathcal{N}$.
Consider a swarm of $m+1$ robots, where $m$ robots lay on $\Q$ and let the last robot, $w$, lay in the safe zone of $\mathcal{N}$.
Let $a$ be the farthest robot from $w$.
Let $b,c$ be the first two found robots, starting from $a$ and following both directions on the perimeter of the associated polygon, one per each direction taken. 
Assume $dist(b,w) > dist (c,w)$.
The problem requires $a$ to move away from $b$ towards a point $x$ such that 
\emph{(i)} $x$ belongs to the safe zone of $\mathcal{N}$, 
\emph{(ii)} $x$ belongs to the halfplane delimited by the line $\bar{bc}$ that does not contain $a$, and
\emph{(iii)} $x$ must not be on any line passing by the position of $w$ and any other robot on $\Q$.
Note that requests \emph{(i,iii)} are imposed in order to have $x$ visible by every robot.
See \Cref{fig:pseudo}.

\end{problem}

\begin{lemma}\label{lemma:pseudo_notin}
	$\pseudo\notin  \prob{\Ofsta{A}}$.
\end{lemma}
\begin{proof}
\pseudo\ cannot be solved in the \async\ mode, only using internal lights.
Let us consider the problem instance given by \Cref{fig:pseudo} where the pseudo-polygon of the initial configuration is composed of $\frac{n}{2}+3$ vertices, with $n=8$.
Let us assume $b$ is activated for the first time during the movement of $a$, when $a$ is hidden by $c$ (i.e. $b,c,a$ are collinear).
When $b$ looks at its snapshot, it recognizes a feasible initial configuration (it sees a pseudo-polygon with $\frac{n}{2}+2$ robots, and the robot $w$).
According to this configuration, $b$ erroneously elects itself as the robot that has to move away from the pseudo-polygon.
It has no means to understand if $a$ exists or not.
On the other hand, $a$ has no means to know if $b$ has updated its internal light to memorize it is not the elected robot to move.
\end{proof}

\subparagraph*{False election.} The impossibility of solving \pseudo\ in the asynchronous modes with just internal lights derives from a critical issue that is typical of swarms with obstructed visibility.
This critical issue can be described as the \emph{false election} phenomenon.
Such phenomenon can be informally described as follows:
from a stable configuration, the given problem requires the use of a leader election routine to elect the unique robot (the \emph{true leader}) which has to execute a non-null movement to reach the next configuration.
All the other robots have to stay still.
In the \async\ mode, a robot $r$ executes its look step while the true leader is moving and is hidden from $r$.
However, $r$ cannot deduct the presence of the true leader from its snapshot.
So, applying the same leader election routine, $r$ elects itself as the (\emph{false}) leader, thus starting an unrequested movement.

The false election phenomenon must be examined when trying to transpose a \semi\ algorithm in the \async\ mode.
In particular, the use of lights must be considered as a possible method to avoid false elections. 
As we have shown in \Cref{lemma:pseudo_notin} for \pseudo, internal lights are not sufficient to cope with them.
Instead, the next lemma proves that external lights are required (and sufficient) to correctly solve the \pseudo\ problem in the \async\ mode.

\begin{lemma}\label{lemma:pseudo_in}
	$\pseudo\in \left(\prob{\Ooblot{S}} \cap\prob{\Ofcom{A}} \right)$.
\end{lemma}
\begin{proof}	
\pseudo\ is solvable in $\Ooblot{S}$ (i.e. in any synchronous model), since complete visibility is guaranteed at any activation time and all the movements (null and non-null) are univocally determined by geometric conditions.
In fact, each robot can determine $\Q$, the watcher $w$, and the robot $a$ (the farthest from $w$).
The robot $a$ can compute its final destination and move there. 
If a robot is not the farthest from the watcher, or if it sees two robots that are not part of the pseudo-polygon, then it stands still.
	
$\pseudo$ needs at least external lights to be solvable in the \async\ mode.
We show here an algorithm that needs 4 colors: \col{off} (default), \col{on}, \col{a}, \col{b}.
In the first epoch, every robot updates its color according to its role: robot $a$ turns into \col{a}, robot $b$ turns into \col{b}, whereas the remainder turns into \col{on}.
Afterward, let $r$ be an activated robot that sees no \col{off} robots and that notes there is only one robot (the watcher) out of the pseudo-polygon.
Let $V_r$ be the set of colors $r$ can see.
\begin{itemize}[noitemsep]
	\item if $V_r=\{\col{a}, \col{b}, \col{on}\}$, $r$ turns into \col{on} and stays still;
	\item if $V_r=\{\col{a}, \col{on}\}$, $r$ turns into \col{b} and stays still;
	\item if $V_r=\{\col{b}, \col{on}\}$, and if $r$ is the farthest robot from $w$, it turns into \col{a} and starts moving;
	\item if $V_r=\{\col{on}\}$, it means $r$ is \col{b} and stays still (robot $a$ is hidden).
\end{itemize}
If a robot $r$ sees two robots not belonging to the pseudo-polygon, then $r$ does not move (the final configuration is already formed or is about to be formed).
	
\end{proof}

\begin{theorem}\label{th:oblotA}
$$\Ooblot{S} > \Ooblot{A}$$
$$\Ofsta{S} > \Ofsta{A}$$
$$\Ofsta{A} \perp \Ooblot{S}.$$
\end{theorem}

\section{Relation map}
\Cref{tab:relation_map} summarizes the results proved in this work, showing the relations ($>$, $<$, $\perp$, and $\equiv$) that hold between the pairs of models in our opaque framework.
The map shows also which of the six witness problems (\trt\ for \triangleroundtrip, \fff\ for \flipflopflip, \nwc\ for \newcomer, \spi\ for \spinning, \ash\ for \angleshift, \pse\ for \pseudo) have been used to prove such relations. 
For some pairs of models (gray cells), the knowledge about what kind of relation holds is still now incomplete.
E.g. between $\Ofsta{F}$ and $\Ofcom{F}$ two possible relations ($<$ or $\perp$) can exist: so far we have built \newcomer\ as witness problem proving that $\newcomer\in\left(\prob{\Ofcom{F}}\setminus \prob{\Ofsta{F}}\right)$.
To prove the orthogonality relation, we should find a witness problem $B$ such that $B\in\left(\prob{\Ofsta{F}}\setminus \prob{\Ofcom{F}}\right)$.
Instead, to prove the strict dominance relation, we should find that any problem in $\Ofsta{F}$ can be solved also under $\Ofcom{F}$.

For the pairs of models where the relation is unknown in the opaque framework, we have reported the relation holding in the transparent framework in red.

\begin{table}[h!]
\caption{Relation map.}
\begin{center}
\small{
\resizebox{\textwidth}{!}{ 
\renewcommand{\arraystretch}{2}{
\begin{tabular}{|c|c|c|c|c|c|c|c|c|c|c|c|}
\hline
$\Rsh$ &	$\Olumi{F}$ & $\Ofcom{F}$ & $\Ofsta{F}$ & $\Ooblot{F}$ & $\Olumi{S}$ & $\Ofcom{S}$ & $\Ofsta{S}$ & $\Ooblot{S}$ & $\Olumi{A}$ & $\Ofcom{A}$ & $\Ofsta{A}$  \\
\hline
\hline

$\Ooblot{A}$ & \colcellD{\makecell{$<$\\ \trt}} &  \colcellD{\makecell{$<$\\ \trt}} & \colcellD{\makecell{$<$\\ \trt}} & \colcellD{\makecell{$<$\\ \spi}} & \colcellD{\makecell{$<$\\ \trt}} & \colcellD{\makecell{$<$\\ \trt}} & \colcellD{\makecell{$<$\\ \trt}} & \colcellD{\makecell{$<$\\ \pse}} & \colcellD{\makecell{$<$\\ \trt}} & \colcellD{\makecell{$<$\\ \trt}} & \colcellD{\makecell{$<$\\ \trt}}   \\
\cline{1-12}

$\Ofsta{A}$ & \colcellD{\makecell{$<$\\ \nwc}} &  \colcellU{\makecell{$<$ or $\perp$, \theircol{$<$} \\ \nwc,}} & \colcellD{\makecell{$<$ \\ \spi}}  & \colcellO{\makecell{$\perp$\\ \trt, \spi}} & \colcellD{\makecell{$<$\\ \nwc}} & \colcellO{\makecell{$\perp$\\ \nwc, \fff}} & \colcellD{\makecell{$<$\\ \pse}} &  \colcellO{\makecell{$\perp$\\ \pse, \trt}} & \colcellD{\makecell{$<$\\ \nwc}} & \colcellO{\makecell{$\perp$\\ \nwc, \fff}}    \\
\cline{1-11}

$\Ofcom{A}$ & \colcellD{\makecell{$<$\\ \fff}} &  \colcellD{\makecell{$<$\\ \fff}} & \colcellO{\makecell{$\perp$\\ \fff, \nwc}} & \colcellO{\makecell{$\perp$\\ \nwc, \spi}} & \colcellD{\makecell{$<$\\ \fff}} & \colcellU{\makecell{$<$ or $\equiv$, \theircol{$<$}\\ $$}} & \colcellO{\makecell{$\perp$\\ \fff, \nwc}} & \colcellU{\makecell{$>$ or $\perp$, \theircol{$\perp$} \\ \nwc,}} & \colcellD{\makecell{$<$\\ \fff}}    \\
\cline{1-10}

$\Olumi{A}$ & \colcellD{\makecell{$<$\\ \ash}} &   \colcellU{\makecell{$<$ or $\perp$, \theircol{$<$} \\ \ash,}}   & \colcellO{\makecell{$\perp$\\ \ash, \nwc}} &  \colcellO{\makecell{$\perp$\\ \ash, \trt}} & \colcellU{\makecell{$<$ or $\equiv$, \theircol{$\equiv$}\\$$}} & \colcellU{\makecell{$>$ or $\perp$, \theircol{$>$} \\ \fff,}}  &\colcellU{\makecell{$>$ or $\perp$, \theircol{$>$} \\ \nwc,}} & \colcellU{\makecell{$>$ or $\perp$, \theircol{$>$} \\ \trt,}}  \\
\cline{1-9}

$\Ooblot{S}$ & \colcellD{\makecell{$<$\\ \trt}} &  \colcellD{\makecell{$<$\\ \trt}} & \colcellD{\makecell{$<$\\ \trt}} & \colcellD{\makecell{$<$\\ \spi}} & \colcellD{\makecell{$<$\\ \trt}} & \colcellD{\makecell{$<$\\ \trt}} & \colcellD{\makecell{$<$\\ \trt}}    \\
\cline{1-8}

$\Ofsta{S}$ & \colcellD{\makecell{$<$\\ \nwc}} &   \colcellU{\makecell{$<$ or $\perp$, \theircol{$<$}\\ \nwc,}} & \colcellD{\makecell{$<$\\ \spi}} & \colcellO{\makecell{$\perp$\\ \trt, \spi}}& \colcellD{\makecell{$<$\\ \nwc}} & \colcellO{\makecell{$\perp$\\ \nwc, \fff}} \\
\cline{1-7}

$\Ofcom{S}$ & \colcellD{\makecell{$<$\\ \fff}} &  \colcellD{\makecell{$<$\\ \fff}} & \colcellO{\makecell{$\perp$\\ \fff, \nwc}} & \colcellO{\makecell{$\perp$\\ \spi, \nwc}} & \colcellD{\makecell{$<$\\ \fff}}   \\
\cline{1-6}

$\Olumi{S}$ & \colcellD{\makecell{$<$\\ \ash}} &  \colcellU{\makecell{$<$ or $\perp$, \theircol{$<$} \\ \ash,}}  & \colcellO{\makecell{$\perp$\\ \ash, \nwc}} & \colcellO{\makecell{$\perp$\\ \ash, \trt}}  \\
\cline{1-5}

$\Ooblot{F}$ & \colcellD{\makecell{$<$\\ \trt}} &  \colcellD{\makecell{$<$\\ \trt}} & \colcellD{\makecell{$<$\\ \trt}}  \\
\cline{1-4}

$\Ofsta{F}$ & \colcellD{\makecell{$<$\\ \nwc}} &  \colcellU{\makecell{$<$ or $\perp$, \theircol{$<$} \\ \nwc,}}  \\
\cline{1-3}

$\Ofcom{F}$ & \colcellU{\makecell{$<$ or $\equiv$, \theircol{$\equiv$}\\ $$}} \\ 
\cline{1-2}

\end{tabular}\label{tab:relation_map}
}}
}
\end{center}

\end{table}
\section{Conclusions}
We have investigated the computational power of the 12 models of collision-intolerant opaque robots, thus presenting the taxonomy of the problems solved in such framework. 
We have taken inspiration from \cite{BFK21, DFPSY16,FSSW23, FSW20} where the authors provide the complete map of the relations held by the same 12 models but considering collision-tolerant transparent robots.

Thus far, the relations proved here in our opaque framework are the same as in the corresponding transparent framework.
The natural question that arises from this observation is whether the relation map of the opaque models is completely identical to the relation map of the transparent models.
To answer this question, future works should find the missing relations among the twelve opaque models in order to obtain the complete hierarchy in the opaque framework.
Among the others, it is worth mentioning the yet unknown relation between  $\Olumi{S}$ and $\Olumi{A}$.
In the transparent framework, the two models have proved to be computationally equivalent \cite{DFPSY16} through the design of a simulator which, with the help of extra light colors, simulates any \semi\ algorithm in the \async\ mode.
This simulator is not adequate to prove the same relation considering opaque robots, precisely because of their obstructed visibility. 
With the \pseudo\ problem, we have presented the \emph{false election} phenomenon whose formalization and investigation will be preparatory to answer this interesting open question: is it possible to simulate a $\Olumi{S}$ algorithm in the \async\ mode, thus proving that $\Olumi{S}$ and $\Olumi{A}$ are two equivalent models also in the opaque framework?
Are constant-size lights sufficient to always avoid the phenomenon of false elections?
In addition, it would be necessary to formalize and study all the critical issues caused by obstructed visibility: such formalizations may be essential for the correct investigation of the missing relations.

\bibliography{complexity_opaque_robots}

\appendix

\section{Proofs of theorems}\label{appendix}
The proofs of the following theorems hold combining the previously stated lemmas and by transitivity.
We use the compacted notation $\{\opaque{X}_1,\dots,\opaque{X}_m\}^{Y_1,\dots,Y_h}$ to indicate all the models in $\{\opaque{X}_i^{Y_j}\}_{\substack{1\leq i \leq m\\1\leq j \leq h}}$ where $X_i\in\{\oblot, \fsta, \fcom, \lumi\}$ and $Y_j\in \{\ftt, \stt, \att\}$.

\begin{customthm}{\ref{th:oblot_weak}}
Given the schedulers $Y_1= \ftt$, $Y_2= \stt$, $Y_3= \att$, it holds
$$\opaque{\fsta}^{Y_i} > \opaque{\oblot}^{\{Y_j\}_{j\geq i}}$$
$$\opaque{\fcom}^{Y_i} > \opaque{\oblot}^{\{Y_j\}_{j\geq i}}$$
$$\opaque{\lumi}^{Y_i} > \opaque{\oblot}^{\{Y_j\}_{j\geq i}}.$$

\end{customthm}
\begin{proof}
$\triangleroundtrip$ cannot be solved under $\Ooblot{F,S,A}$ (by \Cref{lemma:triangle_notin}) but it can be solved under $\{\Ofsta{}, \Ofcom{}, \Olumi{}\}^{\att,\stt,\ftt}$ (by \Cref{lemma:triangle_in}).
Combining the results, we obtain that $\Ooblot{}$ is strictly dominated by $\Ofsta{}$ and $\Ofcom{}$ for a given synchronization mode $Y_i\in \{\ftt, \stt, \att\}$.
The other strict dominances are derived by transitivity.
\end{proof}

\begin{customthm}{\ref{th:lumi_strong}}
$$\Olumi{A} > \Ofcom{A}$$
$$\Olumi{S} > \Ofcom{S,A}$$
$$\Olumi{F} > \Ofcom{S,A}$$
$$\Ofcom{F} > \Ofcom{S,A}.$$
\end{customthm}
\begin{proof}
\flipflopflip\ is solved under $\Ofcom{F}$ and $\Olumi{A,S,F}$ (by \Cref{lemma:flipflopflip_in}) but it cannot be solved under $\Ofcom{S,A}$ (by \Cref{lemma:flipflopflip_notin}).
Combining the results, the strict dominance relations follow.

\end{proof}

\begin{customthm}{\ref{th:fsta_lumi}}
Given the schedulers $Y_1= \ftt$, $Y_2= \stt$, $Y_3= \att$, it holds
$$\opaque{\lumi}^{Y_i} > \opaque{\fsta}^{\{Y_j\}_{j\geq i}}.$$
\end{customthm}
\begin{proof}
By \Cref{lemma:newcomer_in}, \newcomer\ is solved under $\Olumi{A,S,F}$.
By \Cref{lemma:newcomer_notin}, \newcomer\ cannot be solved under $\Ofsta{F,S,A}$.
Combining the results, the strict dominance relations follow.
\end{proof}

\begin{customthm}{\ref{th:fsta_fcom}}
$$\Ofsta{F,S,A} \perp \Ofcom{S,A}.$$
\end{customthm}
\begin{proof}
By \Cref{lemma:flipflopflip_in} and \Cref{lemma:flipflopflip_notin}, \flipflopflip\ is solved in $\Ofsta{F,S,A}$ but not in $\Ofcom{S,A}$.
By \Cref{lemma:newcomer_in} and \Cref{lemma:newcomer_notin}, \newcomer\ is solved in $\Ofcom{S,A}$ but not in $\Ofsta{F,S,A}$.
Combining the results, the orthogonality relations follow.
\end{proof}

\begin{customthm}{\ref{th:fully_strong}}
$$\Ooblot{F} > \Ooblot{S,A}$$
$$\Ofsta{F} > \Ofsta{S,A}$$
$$\Ofcom{F} > \Ofcom{S,A}$$
$$\Ooblot{F} \perp \Ofcom{S,A}$$
$$\Ooblot{F} \perp \Ofsta{S,A}.$$
\end{customthm}
\begin{proof}
The above relations hold combining the previous lemmas and by transitivity:
\begin{itemize}
\item the strict dominance of $\opaque{X}^\ftt$ over $\opaque{X}^{\stt,\att}$ derives from \Cref{lemma:spinning_in} and \Cref{lemma:spinning_notin}, for each $X\in\{\oblot, \fsta, \fcom\}$.
In fact, \spinning\ is solved in $\{\Ooblot{}, \Ofsta{}, \Ofcom{}\}^{\ftt}$ but it is not solved in $\{\Ooblot{}, \Ofsta{}, \Ofcom{}\}^{\stt,\att}$;
\item the orthogonality between $\Ooblot{F}$ over $\Ofcom{S,A}$ holds since \spinning\ is solved in $\Ooblot{F}$ but not in $\Ofcom{S,A}$, and since \newcomer\ is solved in $\Ofcom{S,A}$ but not in $\Ooblot{F}$ (by \Cref{lemma:newcomer_in}, \Cref{lemma:newcomer_notin});
\item the orthogonality between $\Ooblot{F}$ over $\Ofsta{S,A}$ holds since \spinning\ is solved in $\Ooblot{F}$ but not in $\Ofsta{S,A}$, and since \triangleroundtrip\ is solved in $\Ofsta{S,A}$ but not in $\Ooblot{F}$ (by \Cref{lemma:triangle_in}, \Cref{lemma:triangle_notin}).
\end{itemize}
\end{proof}

\begin{customthm}{\ref{th_lumif}}
$$\Olumi{F} > \Olumi{S,A}$$
$$\Ooblot{F} \perp \Olumi{S,A}$$
$$\Ofsta{F} \perp \Olumi{S,A}.$$
\end{customthm}
\begin{proof}
The above relations hold combining the previous lemmas and by transitivity:
\begin{itemize}
\item the strict dominance of $\Olumi{F}$ over $\Olumi{S,A}$ straightforwardly derives from \Cref{lemma:angleshift_innotin}.
In fact, \angleshift\ is solved in $\Olumi{F}$ but it is not solved in $\Olumi{\stt,\att}$;

\item the orthogonality between $\Ooblot{F}$ over $\Olumi{S,A}$ holds since \angleshift\ is solved in $\Ooblot{F}$ but not in $\Olumi{S,A}$, and since \triangleroundtrip\ is solved in $\Olumi{S,A}$ but not in $\Ooblot{F}$ (by \Cref{lemma:triangle_in}, \Cref{lemma:triangle_notin});

\item the orthogonality between $\Ofsta{F}$ over $\Olumi{S,A}$ holds since \angleshift\ is solved in $\Ofsta{F}$ but not in $\Olumi{S,A}$, and since \newcomer\ is solved in $\Olumi{S,A}$ but not in $\Ofsta{F}$ (by \Cref{lemma:newcomer_in}, \Cref{lemma:newcomer_notin}).
\end{itemize}
\end{proof}

\begin{customthm}{\ref{th:oblotA}}
$$\Ooblot{S} > \Ooblot{A}$$
$$\Ofsta{S} > \Ofsta{A}$$
$$\Ofsta{A} \perp \Ooblot{S}.$$
\end{customthm}
\begin{proof}
The above relations hold combining the previous lemmas and by transitivity:
\begin{itemize}
\item for each $X\in\{\oblot, \fsta\}$, $\opaque{X}^{\stt}$ strictly dominates $\opaque{X}^{\att}$  since \pseudo\ can be solved in $\opaque{X}^{\stt}$ but not in $\opaque{X}^{\att}$ (by \Cref{lemma:pseudo_in} and \Cref{lemma:pseudo_notin});

\item the orthogonality between $\Ofsta{A}$ and $\Ooblot{S}$ holds since \pseudo\ is solved in $\Ooblot{S}$ but not in $\Ofsta{A}$, and since \triangleroundtrip\ is solved in $\Ofsta{A}$ but not in $\Ooblot{S}$ (by \Cref{lemma:triangle_in} and \Cref{lemma:triangle_notin}).
\end{itemize}
\end{proof}

\end{document}